\documentclass{v16nufact}

\newcommand{\Photo}{}

\begin{document}
\vspace*{4cm}
\title{ISOSPIN DECOMPOSITION OF THE $\gamma^{(*)} N \to N^*$ TRANSITIONS AS INPUT FOR
CONSTRUCTING MODELS OF NEUTRINO-INDUCED REACTIONS IN THE NUCLEON RESONANCE REGION}

\author{H. KAMANO$^{1,2}$, S. X. NAKAMURA$^3$, T.-S. H. LEE$^4$, and T. SATO$^3$}
\address{
$^1$KEK Theory Center, Institute of Particle and Nuclear Studies (IPNS), 
High Energy Accelerator Research Organization (KEK), Tsukuba, Ibaraki 305-0801, Japan\\
$^2$J-PARC Branch, KEK Theory Center, IPNS, KEK, Tokai, Ibaraki 319-1106, Japan\\
$^3$Department of Physics, Osaka University, Toyonaka, Osaka 560-0043, Japan\\
$^4$Physics Division, Argonne National Laboratory, Argonne, Illinois 60439, USA
}

\maketitle\abstracts{
We present our recent efforts to determine 
the matrix elements associated with the transition between the nucleon and a nucleon resonance
induced by the vector current, which are necessary ingredients
for models of neutrino-induced reactions in the resonance region. 
This is accomplished by making the comprehensive analysis of the data 
for various meson photo- and electro-production reactions off the nucleon
within a sophisticated coupled-channels framework, which is known as 
the ANL-Osaka dynamical coupled-channels model.
We also give a brief introduction to our project for constructing a unified 
neutrino reaction model conducted at the J-PARC Branch of the KEK Theory Center.
}

\section{Introduction}

Constructing an accurate model for neutrino-induced reactions in the nucleon resonance 
($N^*$) region
assumes more importance in making a precise interpretation of the data 
measured in the future neutrino-oscillation experiments.
One of the major unknown parts of the neutrino-induced reactions in the $N^*$ region
is the weak-current matrix elements for the transition from the nucleon to a nucleon resonance.
However, the vector part of the weak-current matrix elements can be determined with 
the high-statistics data of meson photo- and electro-production reactions off the nucleon 
that are provided from the facilities such as JLab, ELSA, MAMI, SPring-8, and ELPH.
In this contribution, we will present our recent efforts to determine 
such matrix elements within a sophisticated coupled-channels framework known as
the ANL-Osaka dynamical coupled-channels (DCC) model.

In Sec.~\ref{sec:2}, we first explain the background and our motivation 
for studying the neutrino-induced reactions, together with a brief introduction to 
our research project for constructing a unified neutrino-nucleus reaction model 
conducted at the J-PARC Branch of KEK Theory Center~\cite{jparc}.
We then present in Sec.~\ref{sec:3} our recent efforts for determining the $\gamma^{(*)} N \to N^*$ 
transition form factors and making its isospin decomposition through 
the comprehensive analysis of the electron-, photon-, and pion-induced meson-production 
reactions off the nucleon within the ANL-Osaka DCC model.
Summary is given in Sec.~\ref{sec:4}.

\section{Motivations for studying neutrino-induced reactions}
\label{sec:2}

Everyone has their own interests in the neutrino-induced reactions. 
However, they may be roughly classified into two categories.
One is to obtain accurate knowledge of neutrino-reaction cross sections.
Such knowledge is necessary for improving neutrino event generators (see, e.g., Ref.~\cite{hayato}), 
and thus it is important for extracting neutrino parameters, 
such as the individual neutrino masses and the leptonic $CP$ phase, 
and searching for physics beyond the standard model
from the future neutrino-oscillation experiments.
Also, precise knowledge of low-energy neutrino-reaction cross sections 
for the nucleon and light-nucleus targets is a key to understanding 
the physics of core-collapse supernovae (see, e.g., Ref.~\cite{nasu}).
This category is therefore related more closely to the particle physics and astrophysics.
Another category comes from the fact that the neutrino reactions can be 
a new source of information on the substructure of the nucleon and baryon resonances, and nuclei.
In fact, the matrix elements associated with the axial transition induced by the weak interaction provide 
the information totally independent from that obtained by electromagnetic probes.
The second category is thus related more closely to the hadron and nuclear physics.
Of course accurate determination of the transition matrix elements requires
the development of reliable models that describe the neutrino-induced reactions very well,
and developing such models can help the studies belonging to the first category, 
particularly those relevant to the accelerator and atmospheric neutrino-oscillation experiments.

\begin{figure}
\centerline{\includegraphics[width=0.5\textwidth,clip]{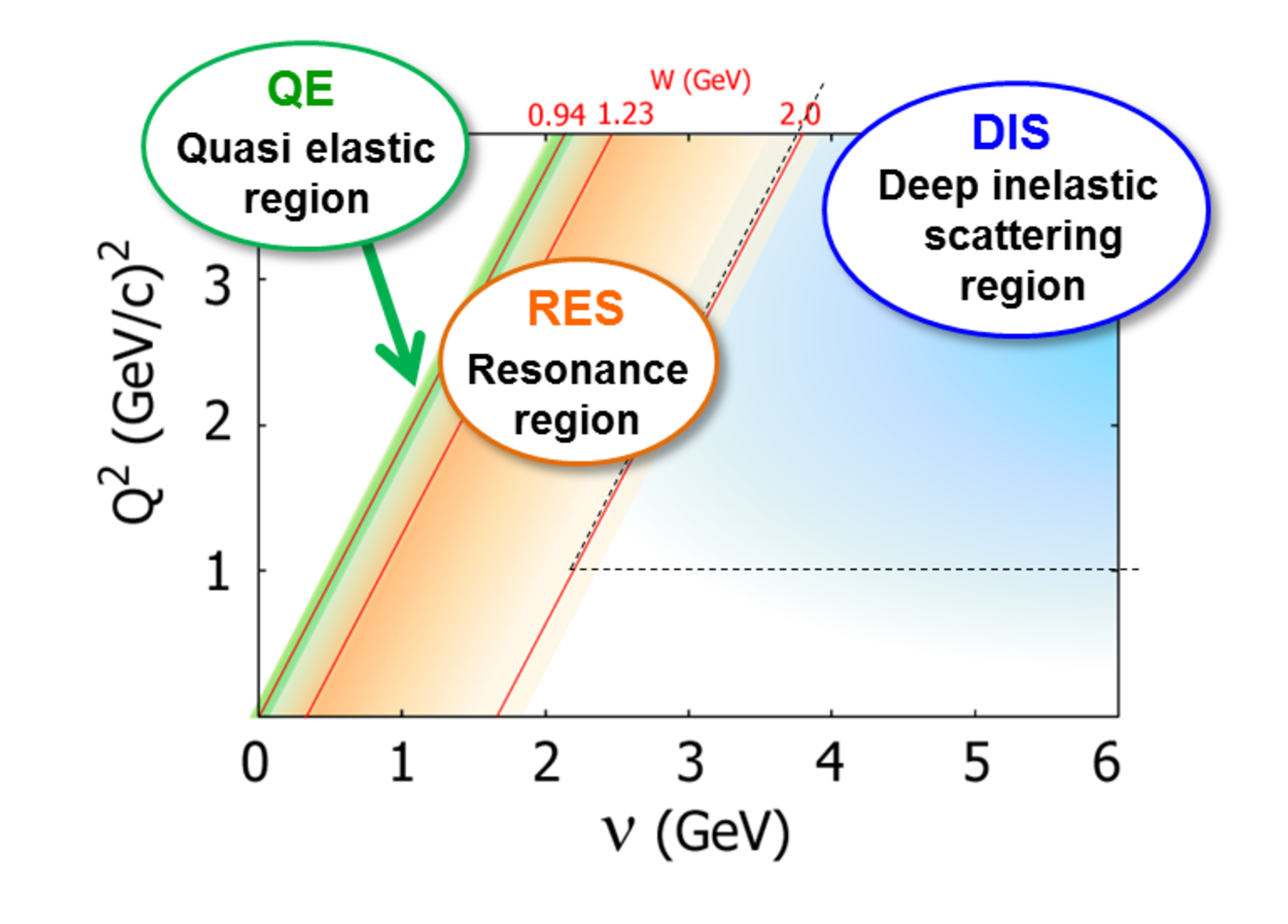}}
\caption[]{
Kinematical region for the neutrino-nucleus reactions relevant 
to the determination of neutrino parameters such as individual neutrino 
masses and leptonic $CP$ phase (the figure is from Ref.~\cite{kamano}).
}
\label{fig:kinematics}
\end{figure}
The neutrino energies relevant to the oscillation experiments are 
from hundreds of MeV to tens of GeV. 
This extends over a quite large kinematical region including the quasi-elastic scattering (QE), 
resonance productions (RES), and deep inelastic scattering (DIS), see Fig.~\ref{fig:kinematics}.
To study the neutrino reactions over such a large kinematical region, one needs to develop 
a reaction model not only for studying each of the three regions individually, 
but also for describing all of them comprehensively including their overlapping regions.
However, the situation is not so simple because each region is governed 
by rather different physics mechanisms and theoretical foundations.
To tackle this challenging issue, theorists and experimentalists from different area of expertise have to get together, 
and this led us to the development of a new collaboration at the J-PARC Branch of KEK Theory Center~\cite{jparc}.
The ultimate goal of this collaboration is to deepen our knowledge of the neutrino-nucleus interaction
by constructing a unified model that comprehensively describes the neutrino-nucleus reactions over 
the QE, RES, and DIS regions.
We quote Ref.~\cite{neutrino-rev} for a detailed review of the activities of our collaboration.
To make this collaboration succeed, of course it is essential to first develop a reliable model in each kinematical region, 
and we are developing a model in the RES region.
Our goal in the RES region is to develop a microscopic model that accurately describes various 
meson productions from neutrino-nucleus reactions exclusively.
However, before going to reactions for nuclear targets, we have to first develop a model at the nucleon level.

\begin{figure}
\centerline{\includegraphics[width=\textwidth,clip]{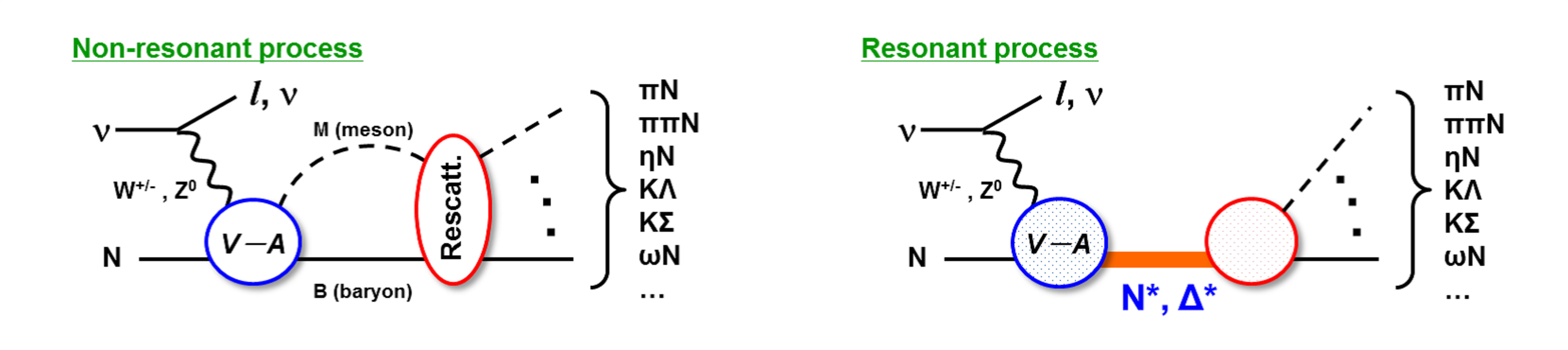}}
\caption[]{
Schematic representation of
the neutrino-nucleon reaction in the RES region.
(Left) Non-resonant process.
(Right) Resonant process.
See the text for the details.
}
\label{fig:process}
\end{figure}
The neutrino-nucleon reaction in the RES region is characterized 
by the so-called nonresonant and resonant processes, which are schematically 
given by the diagrams presented in Fig.~\ref{fig:process}.
Here the blue circles represent the transition matrix element induced 
by the weak interactions, while the red circles represent the rescattering 
process and resonance decay vertex induced by the purely hadronic interactions, respectively.
At very low $Q^2$, the matrix elements for the axial current can be evaluated from $\pi N$ reactions 
and the PCAC hypothesis, while at higher $Q^2$
the neutrino data are definitely needed to determine them.
However, the other ingredients of the neutrino-nucleon reactions can be determined with 
the data of pion-, photon- and electron-induced reactions.
In fact, the hadronic interaction parts (the red circles in Fig.~\ref{fig:process}) 
and the mass and width of the $N^*$ and $\Delta^*$ resonances (the orange thick line
in the resonant process) can be accurately determined 
by analyzing the high-statistics data of these pion- and photon-induced 
reactions off the proton target.
In addition, the matrix elements for the vector current can be determined 
by analyzing the photon- and electron-induced reactions off both the proton and neutron targets.
From now, we focus on presenting our efforts for determining vector current matrix elements 
for the transition between the nucleon and a nucleon resonance
based on the ANL-Osaka DCC model.

\section{Determining the $\gamma^{(*)} N \to N^*$ transition form factors within the ANL-Osaka DCC model}
\label{sec:3}

The basic formula of our model~\cite{msl,knls13}
is the coupled-channels integral equation 
for the partial-wave amplitudes specified by the total angular momentum ($J$),
parity ($P$), and total isospin ($I$):
\begin{equation}
\color{black}
T^{(J^P I)}_{b,a} (p_b,p_a;W) = 
V^{(J^PI)}_{b,a} (p_b,p_a;W)
+\sum_c \int_C dp_c p_c^2 V^{(J^PI)}_{b,c} (p_b,p_c;W) G_c(q;W) T^{(J^PI)}_{c,a} (p_c,p_a;W).
\label{eq:teq}
\end{equation}
Here, the subscripts denote the reaction channels,
$a,b,c= \pi N, \eta N, \pi\Delta, \rho N, \sigma N, K\Lambda, K\Sigma$,
where the $\pi \Delta$, $\rho N$, and $\sigma N$ 
are the quasi-two-body channels that subsequently decay into the three-body $\pi\pi N$ channel
(the reaction channels included are the most relevant ones in the considered energy region);
$p_a$ is the magnitude of the momentum in channel $a$ in the center-of-mass system;
$V$ is the transition potential consisting of hadron-exchange diagrams; 
and $G$ is the Green's function.
By solving Eq.~(\ref{eq:teq}), we can sum up all possible transition processes 
between the reaction channels, and this ensures the two-body as well as 
three-body unitarity in the resulting amplitudes.
The model contains parameters such as cutoff factors and bare coupling constants 
and those are determined by fitting to the existing data of meson production reactions.

Our latest published model~\cite{knls16} was constructed by an extensive 
fit to the data of $\pi N$ and $\gamma N$ reactions as listed here.
The data contains not only unpolarized differential cross sections 
but also spin polarization observables, and this results in fitting 
more than 27,000 data points.
As a result, this model covers the meson production reactions
in the energy range up to $\sim 2$~GeV in the photon lab energy.

\begin{table}[t]
\centering
\caption{\label{tab:proton}
Helicity amplitudes for the $\gamma p \to N^*$ transition at the photon point.
$A^p_{1/2,3/2} = A e^{i\phi}$ ($-90^\circ < \phi < 90^\circ$).
The unit of $A$ and $\phi$ are $10^{-3}$GeV$^{-1/2}$ and degree, respectively.
See the text for the details.
}
\vspace{0.4cm}
{\small
\begin{tabular}{lrrrrrrrr}
\hline
& \multicolumn{4}{c}{$A^p_{1/2}$} & \multicolumn{4}{c}{$A^p_{3/2}$}  
\\
& \multicolumn{2}{c}{Ours~\cite{knls16}}
& \multicolumn{2}{c}{BoGa~\cite{boga-p}}
& \multicolumn{2}{c}{Ours~\cite{knls16}}
& \multicolumn{2}{c}{BoGa~\cite{boga-p}}
\\
\cline{2-3} \cline{4-5} \cline{6-7} \cline{8-9}
Particle $J^P(L_{2I2J})$
&$A$&$\phi$ &$A$&$\phi$ &$A$&$\phi$ &$A$&$\phi$ 
\\
\hline
$N(1535) 1/2^-(S_{11})$    &$  161 $&$   8  $&$ 116\pm  10 $&$   7\pm   6 $&   -    &    -   &-             &-      \\ 
$N(1650) 1/2^-(S_{11})^\dag$    &$   36 $&$ -28  $&$  33\pm   7 $&$ - 9\pm  15 $&   -    &    -   &-             &-      \\ 
$N(1440) 1/2^+(P_{11})^\dag$    &$ - 40 $&$  -9  $&$ -44\pm   7 $&$ -38\pm   5 $&   -    &    -   &-             &-      \\ 
$N(1710) 1/2^+(P_{11})^\dag$    &$ - 47 $&$ -24  $&$  55\pm  18 $&$ -10\pm  65 $&   -    &    -   &-             &-      \\ 
$N(1720) 3/2^+(P_{13})$    &$  131 $&$   7  $&$ 110\pm  45 $&$   0\pm  40 $&$ -34  $&$   12 $&$-150\pm  35 $&$  65\pm  35 $\\ 
$N(1520) 3/2^-(D_{13})^\dag$    &$ - 28 $&$   0  $&$ -21\pm   4 $&$   0\pm   5 $&$  101 $&$    4 $&$ 132\pm   9 $&$   2\pm   4 $\\ 
$N(1675) 5/2^-(D_{15})$    &$    9 $&$  21  $&$  24\pm   3 $&$ -16\pm   5 $&$   49 $&$ - 12 $&$  26\pm   8 $&$ -19\pm   6 $\\ 
$N(1680) 5/2^+(F_{15})$    &$ - 44 $&$ -11  $&$ -13\pm   4 $&$ -25\pm  22 $&$   60 $&$ -  2 $&$ 134\pm   5 $&$ - 2\pm   4 $\\ 
\\                                                                                           
$\Delta(1620)1/2^-(S_{31})$&$  105 $&$   1  $&$  52\pm   5 $&$  -9\pm   9 $&  -     &   -    &-      &-      \\  	 
$\Delta(1910)1/2^+(P_{31})$&$  -1 $&$  -90  $&$  23\pm   9 $&$  40\pm  90 $&  -     &   -    &-      &-      \\  	 
$\Delta(1232)3/2^+(P_{33})^{\dag\dag}$&$ -133 $&$ -16  $&$-131\pm 3.5 $&$ -19\pm   2 $&$- 257 $&$  - 3 $&$-254\pm 4.5 $&$  -9\pm   1 $\\  	 
$\Delta(1700)3/2^-(D_{33})^\dag$&$  128 $&$  19  $&$ 170\pm  20 $&$  50\pm  15 $&$  120 $&$   47 $&$ 170\pm  25 $&$  45\pm  10 $\\  	 
$\Delta(1905)5/2^+(F_{35})^\dag$&$   37 $&$  - 8 $&$  25\pm   5 $&$ -23\pm  15 $&$  -24 $&$  -81 $&$ -50\pm   4 $&$   0\pm  10 $\\  	 
$\Delta(1950)7/2^+(F_{37})$&$ - 69 $&$  -14 $&$ -72\pm   4 $&$  -7\pm   5 $&$ - 83 $&$    2 $&$ -96\pm   5 $&$  -7\pm   5 $\\  	 
\hline
\end{tabular}
}
\end{table}
Table~\ref{tab:proton} shows the helicity amplitudes for the electromagnetic transitions 
between the proton and a nucleon resonance evaluated at the photon point.
These were extracted from the residues of the $\gamma p$ reaction amplitudes 
at each resonance pole position.
Our results~\cite{knls16} are compared with those obtained from the Bonn-Gatchina group~\cite{}.
We first find that the helicity amplitudes for the $\Delta(1232)3/2^+$ resonance 
(the resonance with double dagger in Table~\ref{tab:proton})
agree almost perfectly between the two analyses. 
This indicates that the photo-coupling strength associated with 
this resonance has now been very well determined.
We also see a good agreement for several resonances 
(the resonances with dagger in Table~\ref{tab:proton}).
However, visible discrepancy is observed for the other resonances, 
particularly for the high-mass resonances.
This implies that more extensive and accurate data for $\gamma p$ reactions 
are needed to eliminate this analysis dependence in the extracted value of helicity amplitudes.
In this regards, the so-called (over-)complete experiments, in which 
all the spin polarization observables are measured (see e.g., Ref.~\cite{shkl}), 
are being performed at JLab, ELSA, and MAMI, and 
the upcoming data will help improve our analysis significantly.

\begin{table}
\centering
\caption{\label{tab:neutron}
Helicity amplitudes for the $\gamma n\to N^*$ transition 
at the photon point $A^n_{1/2,3/2} = A e^{i\phi}$ ($-90^\circ < \phi < 90^\circ$).
The unit of $A$ and $\phi$ are $10^{-3}$GeV$^{-1/2}$ and degree, respectively.
See the text for the details.
}
{\small
\begin{tabular}{lrrrrrrrr}
\hline
& \multicolumn{4}{c}{$A^n_{1/2}$} & \multicolumn{4}{c}{$A^n_{3/2}$} 
\\
& \multicolumn{2}{c}{Ours~\cite{knls16}}
& \multicolumn{2}{c}{BoGa~\cite{boga-n}}
& \multicolumn{2}{c}{Ours~\cite{knls16}}
& \multicolumn{2}{c}{BoGa~\cite{boga-n}}
\\
\cline{2-3}
\cline{4-5}
\cline{6-7}
\cline{8-9}
Particle $J^P(L_{2I2J})$ 
&$\bar A_{1/2}$ &$\phi$& $\bar A_{1/2}$ &$\phi$& $\bar A_{3/2}$ &$\phi$& $\bar A_{3/2}$ &$\phi$
\\
\hline
$N(1535) 1/2^-(S_{11})^\dag$ &$-112 $&$  16 $&$-103 \pm 11 $&$   8 \pm  5 $&-      &-      &-            &-            \\ 
$N(1650) 1/2^-(S_{11})$ &$-  1 $&$- 47 $&$  25 \pm 20 $&$   0 \pm 15 $&-      &-      &-            &-            \\ 
$N(1440) 1/2^+(P_{11})$ &$  95 $&$- 15 $&$  35 \pm 12 $&$  25 \pm 25 $&-      &-      &-            &-            \\ 
$N(1710) 1/2^+(P_{11})$ &$ 195 $&$-  8 $&$- 40 \pm 20 $&$- 30 \pm 25 $&-      &-      &-            &-            \\ 
$N(1720) 3/2^+(P_{13})$ &$- 59 $&$   6 $&$- 80 \pm 50 $&$- 20 \pm 30 $&$- 28 $&$- 19 $&$-140 \pm 65$&$   5 \pm 30$\\ 
$N(1520) 3/2^-(D_{13})^\dag$ &$- 43 $&$-  1 $&$- 49 \pm  8 $&$-  3 \pm  8 $&$-110 $&$   5 $&$-114 \pm 12$&$   1 \pm  3$\\ 
$N(1675) 5/2^-(D_{15})^\dag$ &$- 76 $&$   3 $&$- 61 \pm  7 $&$- 10 \pm  5 $&$- 38 $&$-  4 $&$- 89 \pm 10$&$- 17 \pm  7$\\ 
$N(1680) 5/2^+(F_{15})^\dag$ &$  34 $&$- 12 $&$  33 \pm  6 $&$- 12 \pm  9 $&$- 56 $&$-  4 $&$- 44 \pm  9$&$   8 \pm 10$\\ 
\hline
\end{tabular}
}
\end{table}
The extracted helicity amplitudes for the neutron case are presented in Table~\ref{tab:neutron}.
Here the results only for the $I=1/2$ $N^*$ resonances are presented because 
the proton and neutron form factors are the same for the $I=3/2$ $\Delta^*$ resonances.
We see a good agreement for several resonances (the resonance with dagger), 
for which the errors of the amplitudes assigned in the Bonn-Gatchina analysis 
are relatively small.
However, significant analysis dependence similar to the proton case is observed 
for the other resonances.
To eliminate this analysis dependence, the (over-)complete experiments are being performed 
also for the $\gamma d$ reactions and new data will be available soon 
so that we can use them to improve our analysis.
Combining the proton and neutron transition form factors,
we can decompose the electromagnetic transition form factors into isoscalar and isovector parts
by using the following equations: 
\begin{eqnarray}
A^{\rm iso~vector}_{1/2,3/2}&=& -\frac{2}{\sqrt{3}}\left(A^p_{1/2,3/2} -A^n_{1/2,3/2}\right),\\
A^{\rm iso~scalar}_{1/2,3/2}&=&  \frac{1}{2}\left(A^p_{1/2,3/2} +A^n_{1/2,3/2}\right),
\end{eqnarray}
for the $I=1/2$ $N^*$ resonances, and
\begin{equation}
\color{black}
A^{\rm iso~vector}_{1/2,3/2}= \sqrt{\frac{3}{2}}A^p_{1/2,3/2} = \sqrt{\frac{3}{2}}A^n_{1/2,3/2},
\end{equation}
for the $I=3/2$ $\Delta^*$ resonances.
These can be used as input to neutrino reaction models.

\begin{figure}
\centerline{\includegraphics[width=\textwidth,clip]{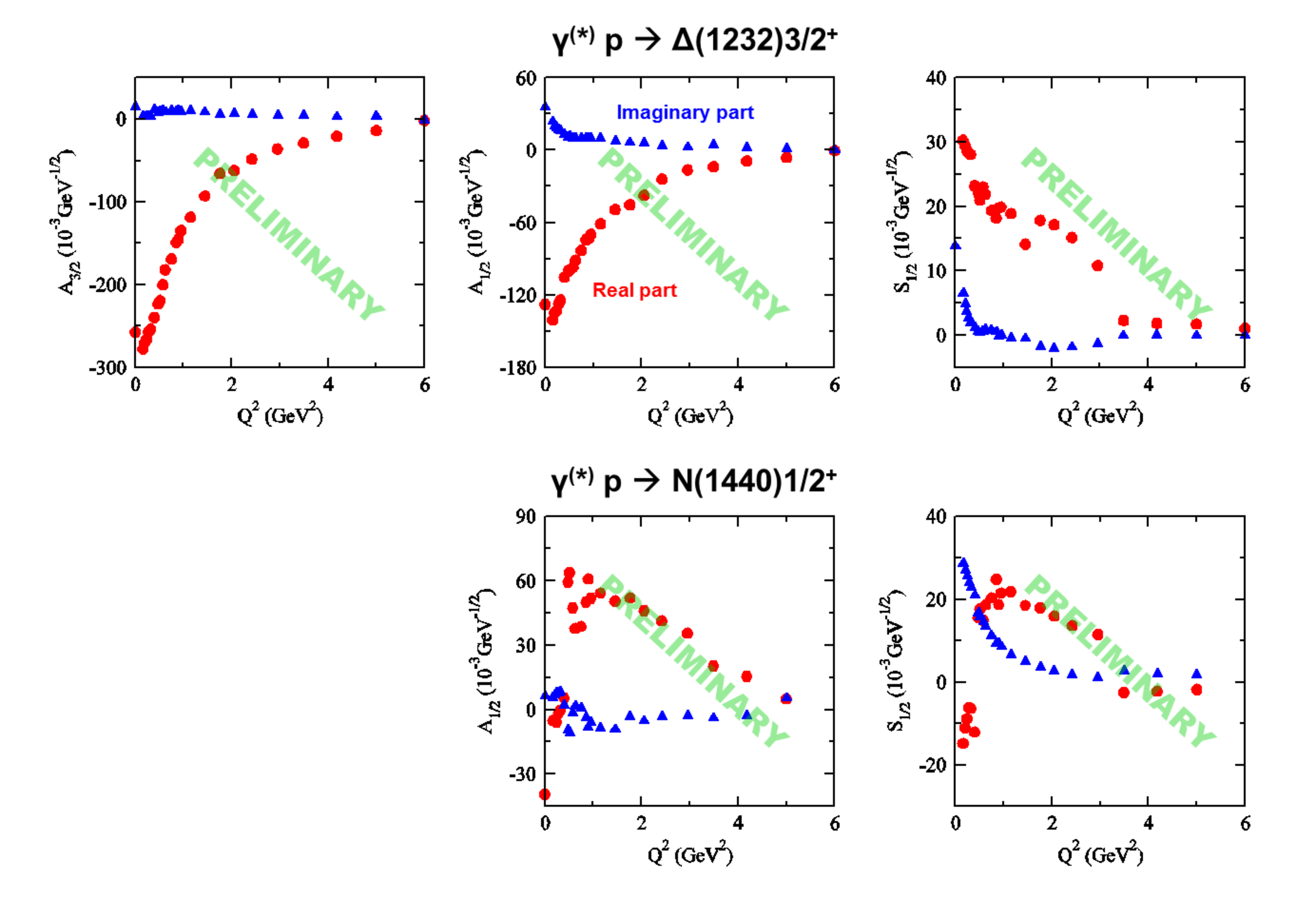}}
\caption[]{
Preliminary results for the helicity amplitudes for the 
$\gamma^* p \to \Delta(1232)3/2^+$ (upper panels)
and $\gamma^* p \to N(1440)1/2^+$ (lower panels) transitions.
The amplitudes are evaluated at the pole positions of the resonances.
Red circles (blue triangles) are the real (imaginary) parts of the amplitudes.
}
\label{fig:emff}
\end{figure}
To obtain the electromagnetic transition form factors at finite $Q^2$, 
one needs to analyze electroproduction reactions.
As a first step, we are now working on analyzing the data for single-pion electroproductions 
off the proton target in the kinematical region up to $Q^2= 6$~GeV$^2$ and $W=1.7$~GeV.
Our electroproduction analysis is ongoing, and in Fig.~\ref{fig:emff} we just present 
a very preliminary results for the $\gamma^* p \to N^*$ transition form factors 
for two resonances, $\Delta(1232)3/2^+$ and $N(1440)1/2^+$.
It is emphasized that the transition form factors are properly defined 
with poles of scattering amplitudes in the complex energy plane.
This may not be so relevant for the purpose of using the form factors 
as input to neutrino reactions, but it is a noticeable progress 
in the field of baryon spectroscopy.
Once our electroproduction analysis is completed, transition form factors 
for other resonances will also be extracted.

\section{Summary}
\label{sec:4}

We have presented our recent effort to determine the electromagnetic transition form factors 
between the nucleon and nucleon resonances, through the DCC analysis of pion-, photon-, and electron-induced 
meson-production reactions off the nucleon.
Our primary motivation for studying transition form factors comes from 
the expectation that those provide crucial information on the quark-gluon substructure of the baryon resonances.
However, those can also be important input for constructing a neutrino reaction model in the 
RES region.
From the comparison with other analysis, we have seen that the extracted helicity amplitudes still 
show visible analysis dependence for several nucleon resonances.
New data for photon- and electron-induced reactions provided from JLab, ELSA, and MAMI
will be used for further improvement of our analysis once they are available.

Finally, we have introduced our research project for constructing 
a unified neutrino reaction model at the J-PARC Branch of KEK Theory Center.
In this regard, we just mention that we have recently developed 
a coupled-channels model for neutrino-nucleon reaction 
in the RES region~\cite{neutrino} by assuming certain $Q^2$ dependence 
for the axial current matrix elements and 
by evaluating vector current matrix elements at finite $Q^2$ 
using the empirical structure functions 
for inclusive electron-neutron reactions.
This will be the basic model for the RES region in our project.

\section*{Acknowledgments}

This work was supported by the JSPS KAKENHI Grant No.~JP25800149 (H.K.) and 
No.~JP16K05354 (T.S.), the MEXT KAKENHI Grant No.~JP25105010 (T.S.),
and by the U.S. Department of Energy, Office of Nuclear Physics Division, 
under Contract No. DE-AC02-06CH11357.

\end{document}